\documentclass[aps,prb,twocolumn,english,superscriptaddress,longbibliography]{revtex4-2}
\usepackage{amsmath,amssymb,amsfonts}
\usepackage{CJK}
\usepackage{bm}
\usepackage{tikz}
\usepackage{babel}
\usepackage{color}
\usepackage{graphicx}

\usepackage{xspace}
\usepackage{epstopdf}
\usepackage{dcolumn}
\usepackage{longtable}
\usepackage{subfigure}
\usepackage{multirow}
\usepackage[colorlinks=true, letterpaper=true, pdfstartview=FitV, linkcolor=blue, citecolor=blue, urlcolor=blue]{hyperref}

\def \H {\mathcal{H}}

\def \S {\mathsf{S}}
\def \P {\mathsf{P}}

\def \T {\mathcal{T}}

\def \H {\mathcal{H}}
\def \K {\hat{\mathcal{K}}}

\def \Z {\mathbb{Z}}

\def \k {\bm{k}}
\usepackage{mathrsfs}

\def \H {\mathcal{H}}

\def \T {\hat{T}}

\def \S {\hat{S}}
\def \P {\hat{P}}
\def \K {\hat{\mathcal{K}}}

\def \Z {\mathbb{Z}}

\def \R {\mathbb{R}}
\def \k {\bm{k}}

\begin{document}
	
	\renewcommand{\figurename}{FIG}
	
	\title{Stability and noncentered $PT$ symmetry  of real topological phases}
	
	\author{S. J. Yue}
	\email[These authors contributed  equally  to this work.]{}
	\affiliation{National Laboratory of Solid State Microstructures and Department of Physics, Nanjing University, Nanjing 210093, China}
	
	\author{Qing Liu}
	\email[These authors contributed  equally  to this work.]{}
	\affiliation{National Laboratory of Solid State Microstructures and Department of Physics, Nanjing University, Nanjing 210093, China}
	
	\author{Shengyuan A. Yang}
	\affiliation{Research Laboratory for Quantum Materials, IAPME, University of Macau, Macau, China}
	
	\author{Y. X. Zhao}
	\email[]{yuxinphy@hku.hk}
	\affiliation{Department of Physics and HK Institute of Quantum Science \& Technology, The University of Hong Kong, Pokfulam Road, Hong Kong, China}

	\begin{abstract}
		Real topological phases protected by the spacetime inversion  ($PT$) symmetry are a current research focus. The basis is that the $PT$ symmetry endows a real structure in momentum space, which leads to $\Z_2$ topological classifications in $1$D and $2$D. Here, we provide solutions to two outstanding problems in the diagnosis of real topology. First, based on the stable equivalence in $K$-theory,
		we clarify that the 2D topological invariant remains well defined in the presence of nontrivial 1D invariant, and we develop a general numerical approach for its evaluation, which was hitherto unavailable.
		Second, under the unit-cell convention, noncentered $PT$ symmetries assume momentum dependence, which violates the presumption in previous methods for computing the topological invariants.
		We clarify the classifications for this case and formulate the invariants by introducing a twisted Wilson-loop operator for both $1$D and $2$D. A simple model on a rectangular lattice is constructed to demonstrate our theory, which can be readily realized using artificial crystals.
	\end{abstract}
	\maketitle

	\section{INTRODUCTION}
	Spacetime inversion ($PT$) symmetry protected topological phases have been attracting increasing interest. The topological classifications of these phases are determined by the $KO$ theory~\cite{zhaoPRL2016a,karoubi2008k}. Particularly, for spinless systems with $(PT)^2=1$, we observe $\Z_2$ classifications  for both $1$D and $2$D gapped systems. The physical origin is that under $PT$ symmetry the wavefunctions over the Brillouin zone (BZ) are essentially \textit{real} in contrast to the usual complex wavefunctions~\cite{Zhao2017}. Hence, the topology should be described by characteristic classes for \textit{real} vector bundles, like the Stiefel-Whitney classes~\cite{nakahara2018geometry}, rather than the Chern classes for complex vector bundles~\cite{TKNN,Haldane_prl,Volovik2013}.
	Under the general framework, various $PT$-symmetric real topological phases have been discovered~\cite{Zhao2017,Sigrist_2017prb,wu2019non,Yang_2018prl,Yang_2019prx,Wang2020}, along with intriguing effects such as non-Abelian braiding structures~\cite{bouhon2020non}, unconventional bulk-boundary correspondence~\cite{Wang2020}, and possibility to switch the spin classes~\cite{Zhao_2021prl}. In terms of physical realizations, graphyne and graphdiyne have been revealed as material candidates for  $2$D and $3$D real topological insulator/semimetal states~\cite{Sheng_19prl,Sheng_2021prb,Sheng_2022prl,Zhu_Phononic,3D_real_Chern}, and the novel nodal-line semimetal with twofold topological charges has been realized by acoustic crystals with projective symmetry~\cite{xue2023stiefel,ZhangBailePRL2015,Zhao_2020prb,Shao_2021prl,Xue_Exper_2022,Qiu_Exper_2022,chen2022brillouin,chen2023classification}.
	
	However, two critical issues remain in the general theoretical framework. First, there is no general method to calculate the 2D topological invariant $w_2$ (also known as the real Chern number), when the 1D (weak) topological invariant $w_1$ is nontrivial. In fact, it was not clear whether $w_2$ is still well defined in such a case. Second,
	in the classification framework, it is standard practice to adopt the unit-cell convention for the Fourier transform~\cite{Kitaev2006,Kitaev2009,chiu2018quantized,Marques_2019prb}, under which the $PT$ operator is presumed to be constant in momentum space. However, this condition is not fulfilled when the $PT$ operation is noncentered~\cite{Inversion_Center_Notes}, i.e., the inversion center not coinciding with the unit-cell center, which are not uncommon in real systems. Then, the conventional Wilson-loop method fails, and the classification itself would be called into question. These two outstanding problems appear as the only obstacles in completing a general theory of $PT$-symmetric real topological phases.

	
	In this Letter, we provide solutions to both problems under the $K$-theoretical framework. For the first
	problem, based on the \textit{stable} equivalence in $K$-theory~\cite{karoubi2008k}, we show that $w_2$ remains well defined in the presence of nontrivial $w_1$, and we develop a working approach to separate $w_2$ from $w_1$ for its evaluation. For the second problem, we justify the invariance of the classification
	for noncentered $PT$ symmetries by the equivariant $K$-theory~\cite{segal1968equivariant} and construct a twisted Wilson-loop operator (tWLO) to formulate the topological invariants. We demonstrate our new approaches in a simple rectangular lattice model, which exhibits interesting topological boundary modes.
	By blowing away the two clouds over the horizon of real topological phases, our work completes the fundamental picture of these phases and provides the theoretical tools to explore them in a much broader scope.

	\section{Separation of weak \& strong topological classifications}
	Let us start by reviewing the meaning of the real topological invariants. As pointed out in Ref.~\cite{Zhao2017}, for a constant $PT$ operator, one can always choose a global basis in momentum space, such that $\hat{P}\hat{T}=\K$ with $\K$ the complex conjugation, and the invariants $w_1$ and $w_2$ can be deduced from the Wilson-loop operator of the \emph{real} valence states. Consider a 2D gapped system and denote the real orthonormal valence states at momentum $\bm k$ as $|\psi_a^{\R}(\k)\rangle$ with $a=1,2,\cdots, \mathcal{N}$ and $\mathcal{N}$ being the number of valence bands. The \emph{real} Wilson-loop operator is defined as
	\begin{equation}\label{D_Wilson}
		D(k_x):= \mathcal{P} \exp\left[\int_{-\pi}^\pi dk_y \mathcal{A}^{\R}_y(k_x,k_y)
		\right]\in O(\mathcal{N}),
	\end{equation}
	where  $[\mathcal{A}^{\R}_{j}(\k)]_{ab}=\langle\psi^{\R}_a(\k)|\partial_{k_j}|\psi^{\R}_b(\k)\rangle$ is the non-Abelian \emph{real} Berry connection, and $\mathcal{P}$ indicates path ordering. We assume that each $|\psi_a^{\R}(\k)\rangle$ is differentiable in the bulk of BZ and periodic along $k_x$.
	
	
	The 1D invariant $w_{1,y}$ is simply defined as
	\begin{equation}\label{w1}
		(-1)^{w_{1,y}}=\det[D(k_x)]=\prod_{a=1}^\mathcal{N} e^{i\theta_a(k_x)},
	\end{equation}
	where $e^{i\theta_a}$ are eigenvalues of $D(k_x)$~\cite{Note_Components}. 
	In either case of $w_{1,y}$, the loop $D(k_x)$ has a further $\mathbb{Z}_2$ topological classification for $\mathcal{N}>2$, i.e., whether the loop can be continuously deformed into a constant loop or not~\cite{Z2_Notes}. This corresponds to the 2D invariant $w_2$~\cite{Zhao2017}.

\begin{figure}
		\centering
		\includegraphics[width=\columnwidth]{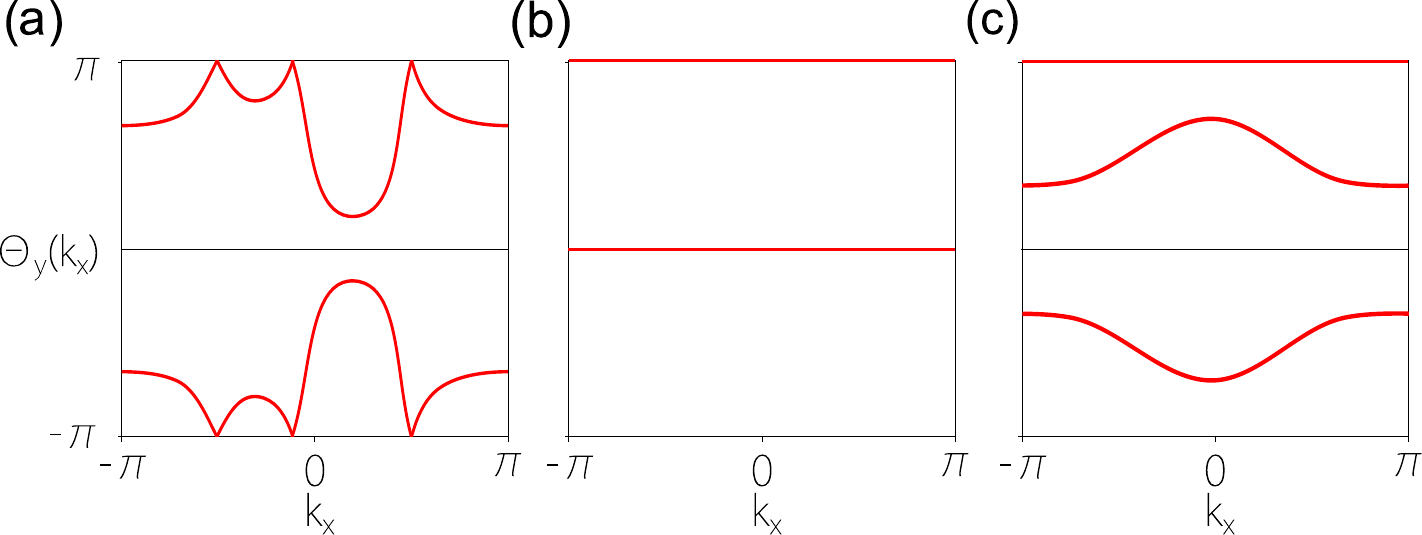}
		\caption{Examples of spectral flows of Wilson-loop operators. (a) A nontrivial $2$-band diagram with $w_{1,y}=0$. (b) The two-band diagram with $w_{1,y}=1$. (c) The generic $3$-band diagram with $w_{1,y}=1$.}
		\label{fig:Theta_Diagrams}
	\end{figure}

	If $w_1=0$, the spectral flow, or more specifically the band structure formed by $\theta_a(k_x)\in [-\pi,\pi)$, offers a way to derive $w_2$. As explained in Ref.~\cite{Yang_2018prl}, $w_2$ is equal to the parity of the number $C_{\pi} $ of linear crossing points on $\theta=\pi$ [see Fig.~\ref{fig:Theta_Diagrams}(a)],
	\begin{equation}\label{w2}
		w_2= C_{\pi} \mod 2.
	\end{equation}
	This is numerically viable, because the spectral flow can be calculated from the \emph{numerical} Wilson-loop operator $W(k_x)$, which, unlike the real Wilson-loop $D(k_x)$, does not require a specific gauge.  For each $k_x$, $W(k_x)$ is related to $D(k_x)$ by a unitary transformation $U(k_x)$~\cite{RuiYu_2011prb,peskin2018introduction,Sheng_2021prb},
	\begin{equation}\label{N_Wilson}
		W(k_x)=U(k_x)D(k_x)U^\dagger(k_x).
	\end{equation}
	Clearly, $U(k)$ is an \textit{uncontrolled} transformation in numerical calculations. Nevertheless, the important point is that $W(k_x)$ and $D(k_x)$ share the same spectrum for each $k_x$, such that $w_2$ can be derived using $W(k_x)$.
	
	For the case of $w_{1,y}=1$, a serious problem arises in applying this numerical method, that there is always a flat $\theta$-band at $\theta=\pi$. For instance, in the case of $\mathcal{N}=2$, the two $\theta$-bands must be flatly fixed at $\theta=0$ and $\pi$, respectively [see Fig.~\ref{fig:Theta_Diagrams}(b)]. For $\mathcal{N}=3$, one $\theta$-band is flatly fixed at $\theta=\pi$, and the other two generically form a complex-conjugate pair [Fig.~\ref{fig:Theta_Diagrams}(c)]. Higher numbers of valence bands follow a similar pattern. Evidently, $w_2$ can no longer be inferred from counting the crossing points at $\pi$~\cite{Patch_Notes}, and it was suspected that $w_2$ may not be well defined for such a case.

	Our solution to the problem is inspired by looking into the fundamental $K$-theoretical classification over the 2D Brillouin torus~\cite{Kitaev2009,zhaoPRL2016a}:
	\begin{equation}\label{Cls-grp}
		\widetilde{KO}(T^2)=(\Z_2\oplus \Z_2)_w\oplus (\Z_2)_s.
	\end{equation}
	Here, the two components in the first bracket correspond to $w_{1,x}$ and $w_{1,y}$, the last component corresponds to $w_2$, and the subscript `$w$' and `$s$' refer to the weak and strong insulators in $2$D, respectively.
	It is significant to observe that from the viewpoint of $K$-theory, the topological structure in $2$D is \emph{separable} from that in $1$D. This has two important consequences. First, $w_2$ must remain well defined regardless of the value of $w_1$. Second, the weak invariants $w_{1,x/y}$ can be canceled out by forming a direct sum with an appropriate Hamiltonian which has trivial $w_2$ but nontrivial $w_{1,x/y}$, so that $w_2$ is preserved under the operation.
	
	The above idea leads to the following working procedure. Let $\H(k_x,k_y)$ be the Hamiltonian of the 2D insulator system with a nontrivial $w_{1,y}$. We construct the direct sum of $\H(k_x,k_y)$ with $\H(0,-k_y)$ to obtain a composite Hamiltonian
	\begin{equation}\label{summed_H}
		\mathbf{H}(\k)=\begin{bmatrix}
			{\H}(k_x,k_y)&\Delta\\
			\Delta^\dagger & {\H}(0,-k_y)
		\end{bmatrix} .
	\end{equation}
	Here, $\Delta$ is a perturbation that couples $\H(k_x,k_y)$ and $\H(0,-k_y)$. It is needed to remove the flat bands at $\theta=\pi$. We must require $\Delta$ to preserve the $PT$ symmetry, namely $[\Delta,\P\T]=0$. For example, we may simply choose it to be $\lambda I$, with $\lambda$ a real number and $I$ the identity matrix.
	Then, as long as $\Delta$ is small enough that does not close the energy gap, $\mathbf{H}(k_x,k_y)$ and $\H(k_x,k_y)$ share the same $w_2$. But now, because $\mathbf{H}(k_x,k_y)$ has a trivial $w_{1,y}$ [due to the cancelation between $\H(k_x,k_y)$ and $\H(0,-k_y)$], we can directly apply the numerical Wilson-loop method on $\mathbf{H}(k_x,k_y)$ to derive this $w_2$.
	
	Clearly, for a system with both nontrivial $w_{1,y}$ and $w_{1,x}$, we just need to further add $\H(-k_x,0)$ into the direct sum, along with $PT$-preserving perturbation terms to couple the diagonal blocks.
	
	\begin{figure}
		\centering
		\includegraphics[width=\columnwidth]{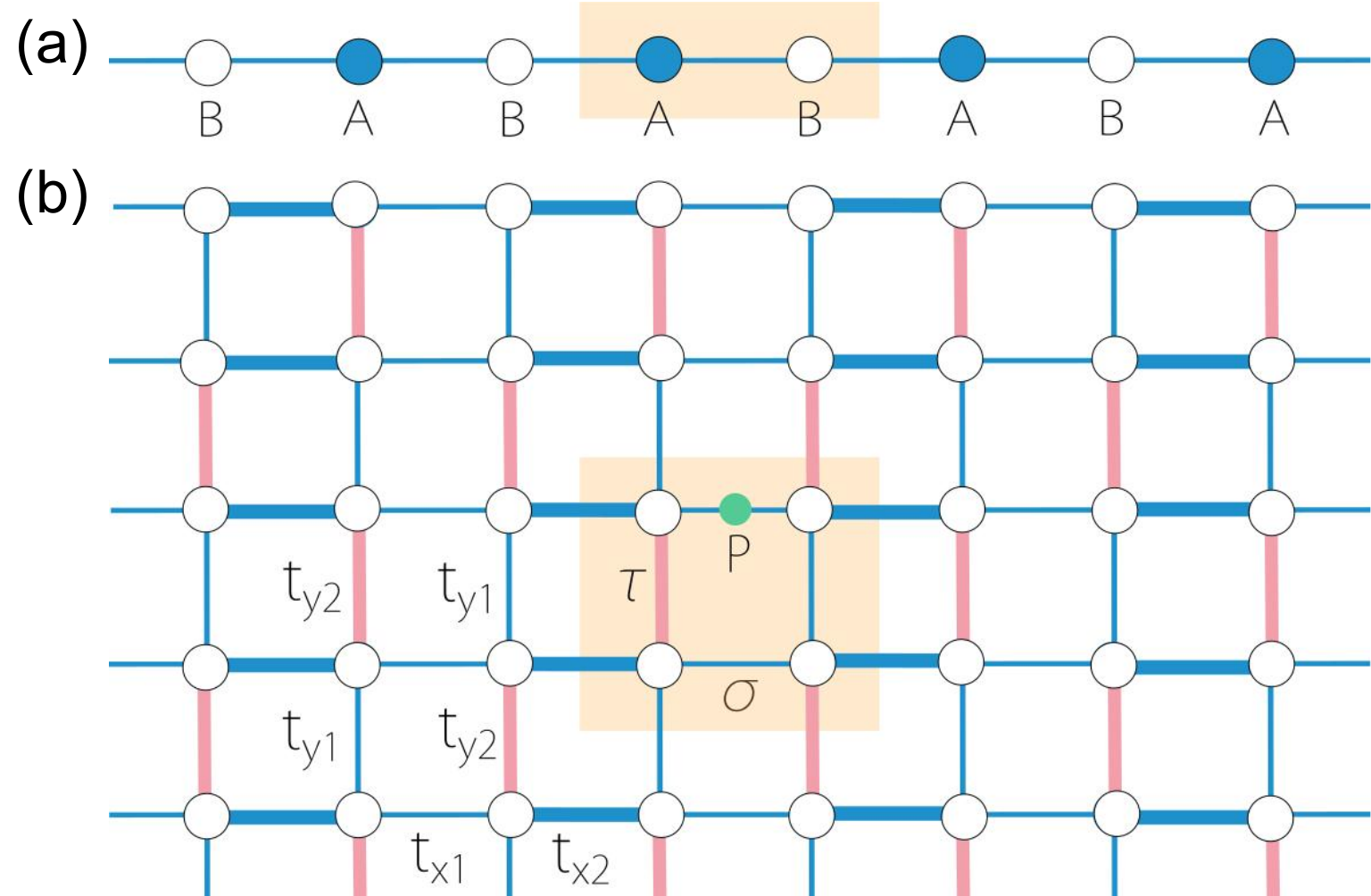}
		\caption{(a) A chain model. A and B sublattices have different on-site energies and all hopping amplitudes are equal. Two inversion centers are marked by blue and red dash lines. (b) A rectangular model. The yellow shadowed unit cell is determined by the dimerization pattern indicated by thick and thin bonds. Negative and positive real hopping amplitudes are marked in red and blue, respectively. }
		\label{fig:lattice}
	\end{figure}
	
	\section{Noncentered $PT$ operators and the twisted Wilson loop}
	In analyzing topological invariants, it is standard practice to adopt the unit-cell convention in the Fourier transform from real space to momentum space~\cite{Kitaev2006,Kitaev2009}. In this convention, only the distances $\bm{R}$ between unit cells
	appear in the Hamiltonian $\H(\k)$, whereas the relative positions of sites within a unit cell are gauged out, such that $\H(\k)$ is manifestly periodic in momentum space.
	The periodicity is compatible with the fact that BZ is topologically a torus, which is essential for defining topological invariants~\cite{Kitaev2009}. 
	
	
	{However, an additional complexity arises when the spatial symmetries are non-centered. Here, we demonstrate and elaborate this complexity by $PT$ symmetry. }
	
	
	{First of all, we stress that in practice, it is often necessary to consider noncentered $PT$ symmetry.  First, there exist wide range of lattices where choosing a primitive unit cell with centered $PT$ is simply impossible~\cite{EvenCell_Notes}.  For example, it is the case if each unit cell contains even number of sites with the inversion center at one site. The 1D chain in Fig.~\ref{fig:lattice}(a) serves as a simple example, and a 2D example investigated in detail with our method can be found in Appendix~\ref{A}. Second, for certain lattices, although it is possible to 
		choose a unit cell with centered $PT$, such a choice may not be compatible with a given boundary geometry that we want to study. For example, to study the 2D lattice in Fig.~\ref{fig:lattice}(b) with a flat edge, the boundary-compatible choices of a cell must have noncentered $PT$. }
	
	{For all these cases, the momentum-space operator $\P\T$  acquires $\k$ dependence under the unit-cell convention~\cite{Inversion_Center_Notes}. This violates the presumption in the topological classification scheme that requires a constant $\P\T$ operator in momentum space~\cite{Zhao2016}. To illustrate this point, let's consider the model in Fig.~\ref{fig:lattice}(b).} The $\hat{P}\hat{T}$ operator takes the following form,
	\begin{equation}\label{PT_1D}
		\hat{P}\T=\begin{bmatrix}
			1 & 0\\
			0 &e^{ik_y}\\
		\end{bmatrix}_\tau \otimes \sigma_1 \K,
	\end{equation}
	where $\tau$ and $\sigma$ are two pseudospins corresponding to the row and the column degrees of freedom of the four sites in a unit cell, and the phase factor $e^{ik_y}$ comes from the fact that $P$ maps the two lower sites in the shadowed unit cell into its upper neighbor, i.e., from the noncentered character of $P$. Note that the fundamental algebra
	\begin{equation}\label{Unit_Square}
		(\hat{P}\T)^2=1
	\end{equation}
	is still respected. Equation \eqref{Unit_Square} indicates that \textit{locally} at each $\k$ we can always choose a basis which corresponds to a real Hamiltonian, i.e., we can strip out the $k$-dependence of $\P\T$, and transform it to be $\P\T=\K$. However, the associated unitary transformation $V(\k)$ must depend on $\k$, and consequently the transformed Hamiltonian $\tilde{\H}(\k)=V(\k)\H(\k)V^\dagger(\k)$ is generally \emph{non-periodic} in momentum space. For Eq.~\eqref{PT_1D}, we have $V(\k)=e^{-i(k_y-\pi)/4}e^{ik_y\tau_3/4}e^{-i\sigma_1\pi/4}$, and the transformed operator $\tilde{P}\tilde{T}=V\P\T V^\dagger=\K$. Although $\tilde{\H}(\k)$ is non-periodic, it satisfies a twisted periodic boundary conditions (tPBCs):
	\begin{equation}\label{B_Cond}
		\tilde{\H}(k_x,k_y+2\pi)=\Lambda \tilde{\H}(k_x,k_y) \Lambda^\dagger,\quad  \Lambda^2=1.
	\end{equation}
	Here, $\Lambda=e^{i\alpha}V(k_y+2\pi)V^\dagger(k_y)$ is in general a real matrix for some appropriately chosen $\alpha$.
	For Eq.~\eqref{PT_1D}, $\Lambda=\tau_3\otimes\sigma_0$~\cite{Cell_periodic_Note}.
	
	We now show that under the tPBCs \eqref{B_Cond}, the topological classification is still given by \eqref{Cls-grp}. Since $\Lambda^2=1$, the period of $\tilde{\H}(\k)$ is $4\pi$ for $k_y$. With the $4\pi$ periodicity, the tPBCs~\eqref{B_Cond} are embodied as a $\mathbb{Z}_2$ group action:
	\begin{equation}\label{Group_Action}
		\lambda:~\tilde{\H}(k_x,k_y) \mapsto \tilde{\H}(k_x,k_y+2\pi)= \Lambda\tilde{\H}(k_x,k_y)\Lambda^\dagger.
	\end{equation}
	Clearly, $\lambda^2=1$. Hence, the topological classification of $\tilde{\H}(\k)$ with $k_y\in [0,2\pi)$ under the tPBCs~\eqref{B_Cond} is equivalent to the classification of $\tilde{\H}(\k)$ with $k_y\in [0,4\pi)$ under the PBCs and the $\mathbb{Z}_2$ group action \eqref{Group_Action}, i.e., the classification is given by the equivariant orthogonal $K$ group $\widetilde{KO}_{\Z_2}(T^2)$, where the torus $T^2$ denotes the doubled BZ with $k_y\in [0,4\pi)$. It is important to note that the $\Z_2$ group action $\lambda$ is \textit{free} on $T^2$, which leads to $\widetilde{KO}_{\Z_2}(T^2)\cong \widetilde{KO}(T^2/{\Z_2}) $~\cite{segal1968equivariant}. Because topologically $T^2/\Z_2\cong T^2$, it turns out that
	\begin{equation}
		\widetilde{KO}_{\Z_2}(T^2)\cong \widetilde{KO}(T^2),
	\end{equation}
	and the classification is still given by \eqref{Cls-grp} as claimed.
	
	
	Although the topological classification is unchanged, the formulation of the topological invariants has to be modified. 
	Here, we introduce the tWLO $W^{(\Lambda)}$ designated for the tPBCs \eqref{B_Cond}, which systematically formulates both $1$D and $2$D invariants.


	Consider a $1$D insulator (sub)system. We equally divide the 1D BZ into $N$ intervals, where the separation points are labeled by $i=0,1,\cdots,N-1$. Recall that the ordinary Wilson-loop operator $W$ is evaluated by
	\begin{equation}
		\begin{aligned}\label{Wilson_loop}
			W=\lim_{N\rightarrow\infty}\prod_{i=0}^{N-1}F_{i,i+1},\\
		\end{aligned}
	\end{equation}
	with $[F_{i,i+1}]_{ab}=\langle a,k_{i}|b,k_{i+1}\rangle$~\cite{RuiYu_2011prb}.
	Here, $a$ and $b$ label the $\mathcal{N}$ valence bands, and therefore $F_{i,i+1}$ is an $\mathcal{N}\times\mathcal{N}$ matrix. Note that $[F_{N-1,N}]_{ab}=\langle a,k_{N-1}|b,k_0\rangle$, which stems from the periodicity in momentum space.
	
	Here, to accomodate the tPBCs \eqref{B_Cond}, we need to modify Eq.~(\ref{Wilson_loop}) by introducing the tWLO as
	\begin{equation}\label{tWL}
		W^{(\Lambda)}=\lim_{N\rightarrow\infty}\left(\prod_{i=0}^{N-2}F_{i,i+1}\right)\times F^{(\Lambda)}_{N-1,N},
	\end{equation}
	with
	\begin{equation}\label{end_F}
		[F^{(\Lambda)}_{N-1,N}]_{ab}=\langle a,k_{N-1}|\Lambda |b,k_{0}\rangle.
	\end{equation}
	The difference from \eqref{Wilson_loop} lies in the last term. By inserting $\Lambda$, $[F^{(\Lambda)}_{N-1,N}]_{ab}$ corresponds to the desired infinitesimal transition amplitudes, because $\Lambda|b,k_0\rangle$ is an eigenvector of $\tilde{\H}(\k_\perp,k_y+2\pi)$, as immediately seen from \eqref{B_Cond}.
	
	To see that the tWLO is well defined, we note that $W^{(\Lambda)}$ transforms under the gauge transformation $|a,k_i\rangle\rightarrow \sum_{b}U^\dagger_{ba}(k_i)|b,k_i\rangle$ as
	\begin{equation}
		W^{(\Lambda)}\rightarrow U(k_0)W^{(\Lambda)}[U(k_0)]^\dagger,
	\end{equation}
	which is exactly the same transformation rule for the ordinary Wilson-loop operator. Moreover, in the basis of real valence eigenstates $|\psi^{\mathbb{R}}_a\rangle$, it is easy to see from \eqref{B_Cond} that the \emph{real} tWLO $D^{(\Lambda)}$ satisfies
	\begin{equation}
		D^{(\Lambda)}\in O(\mathcal{N}),
	\end{equation}
	which resembles Eq.~\eqref{D_Wilson}. In numerical calculations with uncontrolled gauge, we can just use $W^{(\Lambda)}$
	(instead of $W$) to evaluate the topological invariants $w_{1,2}^{(\Lambda)}$, following the same procedure as in \eqref{w1} and \eqref{w2}.

	In the presence of a nontrivial $w_1^{(\Lambda)}$, $w_2^{(\Lambda)}$ can be extracted by our method in Eq.~\eqref{summed_H}, i.e., by forming the direct sum $\mathbf{H}$. It should be noted that the added perturbation term $\Delta$ in \eqref{summed_H} must satisfy the boundary condition
	$
	\mathbf{H}(k_x,k_y+2\pi)=\mathbf{\Lambda}\mathbf{H}(k_x,k_y)\mathbf{\Lambda}
	$, besides the $PT$ symmetry. Thus, we have the following requirements:
	\begin{equation}\label{Correlation_Cond}
		[\Lambda,\Delta]=0,\quad [\tilde{P}\tilde{T},\Delta]=0.
	\end{equation}

	\section{Model and bulk-boundary correspondence}
	\begin{figure}
		\centering
		\includegraphics[scale=0.38]{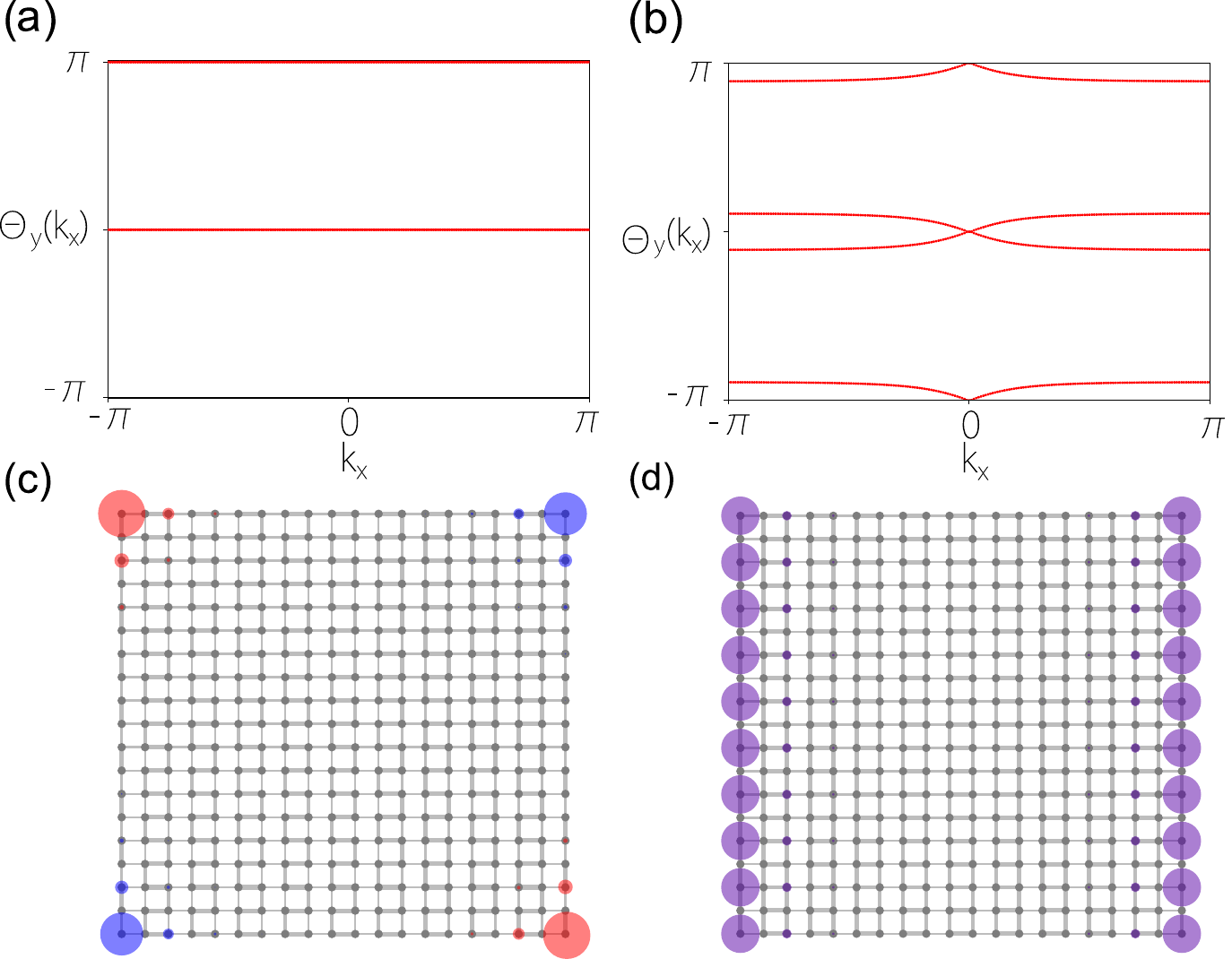}
		\caption{(a) Spectral flow of the tWLO. Here, $t_x^1=1$, $t_x^2=5$, $t_y^1=2$, $t_y^2=4.5$. (b) Spectral flows of the tWLO for $\mathbf{H}$. We choose $\Delta=t_{01}\tau_0\otimes\sigma_1+t_{31}\tau_3\otimes\sigma_1$ with $t_{01}=t_{31}=2$.  (c) The second-order topological phases for a rectangular-shaped sample with $20\times 19$ sites. A $PT$-related pair of zero-energy corner  modes are marked by red (blue) dots on the diagonal (off-diagonal) direction when $J_->0$ ($J_-<0$). (d) $PT$-related first-order helical edge states when $J_-=0$. \label{fig:Model} } 	
	\end{figure}
	{To demonstrate our theory, we present a $2$D insulator model with $(w^{(\Lambda)}_1, w^{(\Lambda)}_2)=(1,1)$. This model can illustrate both our methods, namely to shoot two birds with one stone. More examples can be found in Appendix~\ref{B} and Appendix~\ref{C}, including a $2$D model with centered $PT$ and $(w_1,w_2)=(1,1)$, and a $3$D model of second-order real nodal-line semimetal.}
	
	We note that in contrast to $w_1$ for centered $PT$, $w^{(\Lambda)}_1$ does not lead to boundary states~\cite{chiu2018quantized}. This can be easily understood from the 1D chain in Fig.~\ref{fig:lattice}(a). Here, the two $w^{(\Lambda)}_1$ values determine whether the Wannier center is located at site $A$ or site $B$~\cite{Zak_phase}. Since the Wannier center coincides with a lattice site, the two cases correspond to two distinct atomic insulators. Hence,  $w^{(\Lambda)}_1$ does not have a direct boundary consequence. Boundary states for such systems stem solely from $w^{(\Lambda)}_2$.

	As illustrated in Fig.~\ref{fig:lattice}(b), our model is an appropriately dimerized rectangular lattice with only the nearest neighbor hopping. 
	Under the unit-cell convention, the Hamiltonian is given by
	\begin{equation}\label{Model_H}
		\H(\k)=\sum_{i=1}^{4}f_i(\k)\Gamma^i+g_1(k_y)i\Gamma^4\Gamma^5+g_2(k_y)i\Gamma^3\Gamma^5.
	\end{equation}
	Here, the coefficient functions are given by $f_1=t_x^1+t_x^2\cos k_x$, $f_2=t_x^2\sin k_x$, $f_3=J_+\sin k_y$, $f_4=J_+(1-\cos k_y)$, $g_1=J_-\sin k_y$, and $g_2=J_-(1+\cos k_y)$, where $J_{\pm}=(t_y^1\pm t_y^2)/2$, and the $t$'s are hopping amplitudes as indicated in Fig.~\ref{fig:lattice}. The Dirac matrices are chosen as $\Gamma^1=\tau_0\otimes\sigma_1$, $\Gamma^2=\tau_0\otimes\sigma_2$, $\Gamma^3=\tau_2\otimes\sigma_3$, $\Gamma^4=-\tau_1\otimes\sigma_3$, and $\Gamma^5=-\tau_3\otimes\sigma_3$, which satisfy $\{\Gamma^\mu,\Gamma^\nu\}=2\delta^{\mu\nu}1_4$.

	The noncentered $PT$ operator is given in \eqref{PT_1D}. Under the aforementioned unitary transformation $V(\k)$,
	the Hamiltonian is converted to
	\begin{equation*}\label{Ham_noP}
		\tilde{\H}(\k)=\tilde{f}_1(k_x)\Gamma^1+\tilde{f}_2(k_x)i\Gamma^1\Gamma^2+\tilde{f}_3(k_y)i\Gamma^1\Gamma^3+\tilde{g}(k_y)i\Gamma^3\Gamma^5,
	\end{equation*}
	where $\tilde{f}_1=t_x^1+t_x^2\cos k_x$, $\tilde{f}_2=-t_x^2\sin k_x$, $\tilde{f}_3=-t_y^1\sin (k_y/2)-t_y^2\sin (k_y/2)$, and $\tilde{g}=t_y^1\cos (k_y/2)-t_y^2\cos (k_y/2)$. One can easily check that the tPBC in \eqref{B_Cond} holds, with $\Lambda=\tau_3\otimes \sigma_0$.

	As seen from Fig.~\ref{fig:Model}(a), the tWLO along $k_y$ indicates $w_{1,y}^{(\Lambda)}=1$. Hence, to evaluate $w_{2}^{(\Lambda)}$, we should follow Eq.~\eqref{summed_H} to construct the $\mathbf{H}$ Hamiltonian. The allowed perturbation terms $\Delta$ that satisfy Eq.~\eqref{Correlation_Cond} are real linear combinations of $\tau_0\otimes\sigma_0$, $\tau_0\otimes\sigma_1$, $\tau_0\otimes\sigma_3$, $\tau_3\otimes\sigma_1$, and $\tau_3\otimes\sigma_3$. Then, the spectral flows of the tWLOs for $\mathbf{H}$ is shown in Fig.~\ref{fig:Model}(b). As one can see, the nontrivial strong topology $w_2^{(\Lambda)}$ is resolved from the weak topology.
	
	The strong topology $w_2^{(\Lambda)}=1$ will lead to a one-to-many bulk-boundary correspondence as long as the $PT$ symmetry is preserved~\cite{Wang2020}.
	Particularly, two second-order boundary mode configurations featuring a pair of $PT$-related corner states are shown in Fig.~\ref{fig:Model}(c). Moreover, without closing the bulk energy gap, we observe a first-order case with a pair of $PT$-related helical edge states as the critical state between the two second-order phases [Fig.~\ref{fig:Model}(d)]. 

	\section{Summary and discussion} {
		Based on the $K$-theoretical framework, we resolve two outstanding problems in the theory of real topological phases, which significantly extends the physical relevance and the scope of $PT$-symmetric real topological phases in real materials as well as in artificial systems. Furthermore, the ideas underlying our solutions can be generalized into broader contexts. The method to separate topological invariants at different levels (like $w_1$ and $w_2$) is applicable to all topological classifications based on $K$-theory, and the twisted Wilson loop may be adapted to other momentum-dependent symmetry operators, which are ubiquitous in crystals.}
	
	\appendix
	\renewcommand\thefigure{\Alph{section}\arabic{figure}}   
	\section{ 2D topological insulator with noncentered $PT$ symmetry}\label{A}
	
	It is not true that every $PT$-invariant lattice model has a primitive cell whose center is also the inversion center. There do exist $PT$-invariant lattices for which one can never find a primitive unit cell satisfying the above condition. For example, we can consider a lattice for which the unit cell contains an even number of sites and the inversion center is at one of the lattice site. Suppose one can choose a primitive cell for which the inversion center is also the unit-cell center, then sites in the unit cell form inversion-related pairs except the one at the center. This leads to an odd number of sites, which is a contradiction. For this class of lattices, one has to adopt our method to study the real topology. 
	\setcounter{figure}{0}
	\begin{figure}
		\centering	\includegraphics[scale=0.25]{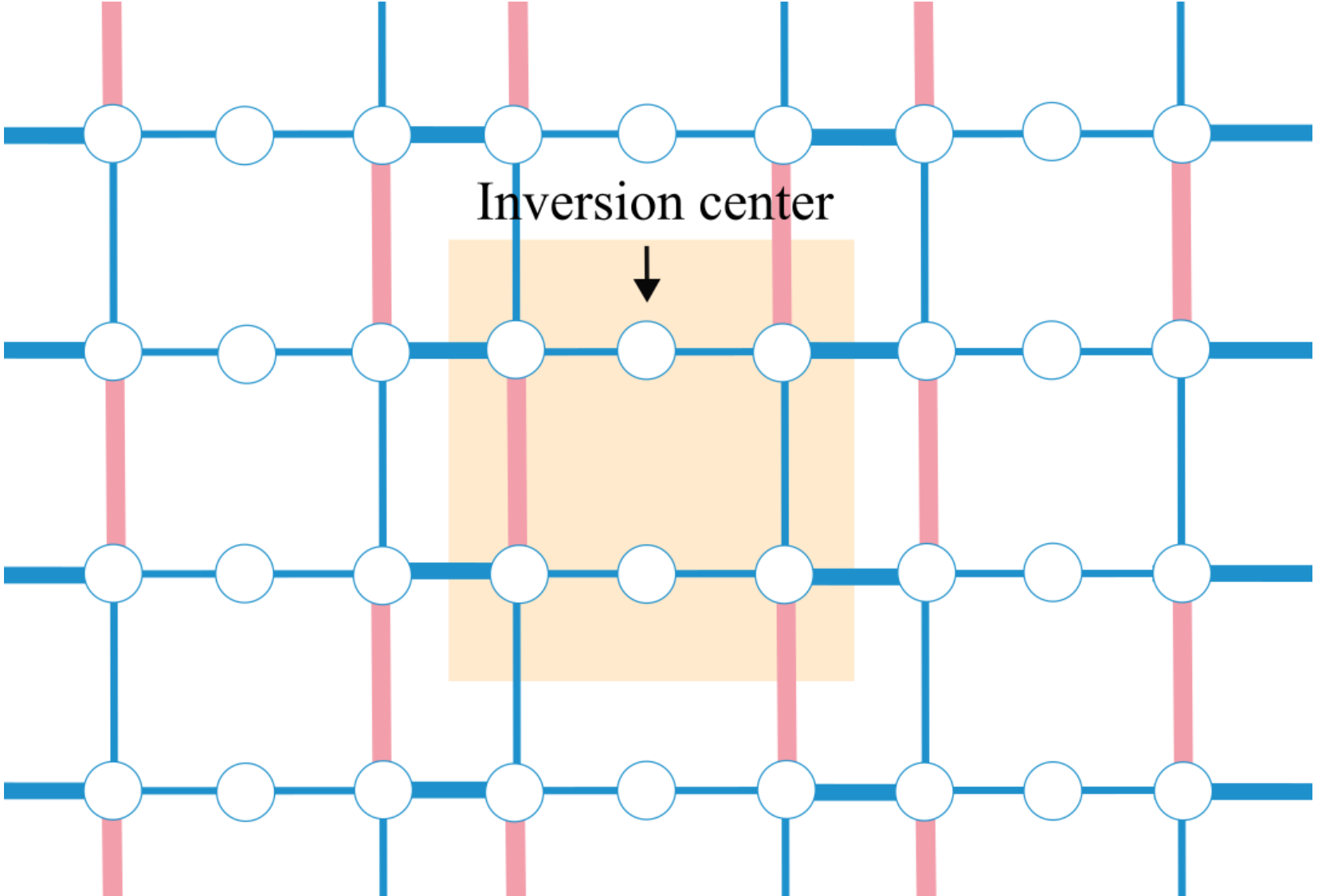}
		\caption{A 2D lattice model. The thickness of each bond corresponds to the magnitude of the hopping amplitude, and negative and positive real hopping amplitudes are marked in pink and blue, respectively. A unit cell is marked by the shadowed rectangle, which contains six sites. }
		\label{model_3}
	\end{figure}
	
	\begin{figure}
		\centering	
		\includegraphics[scale=0.45]{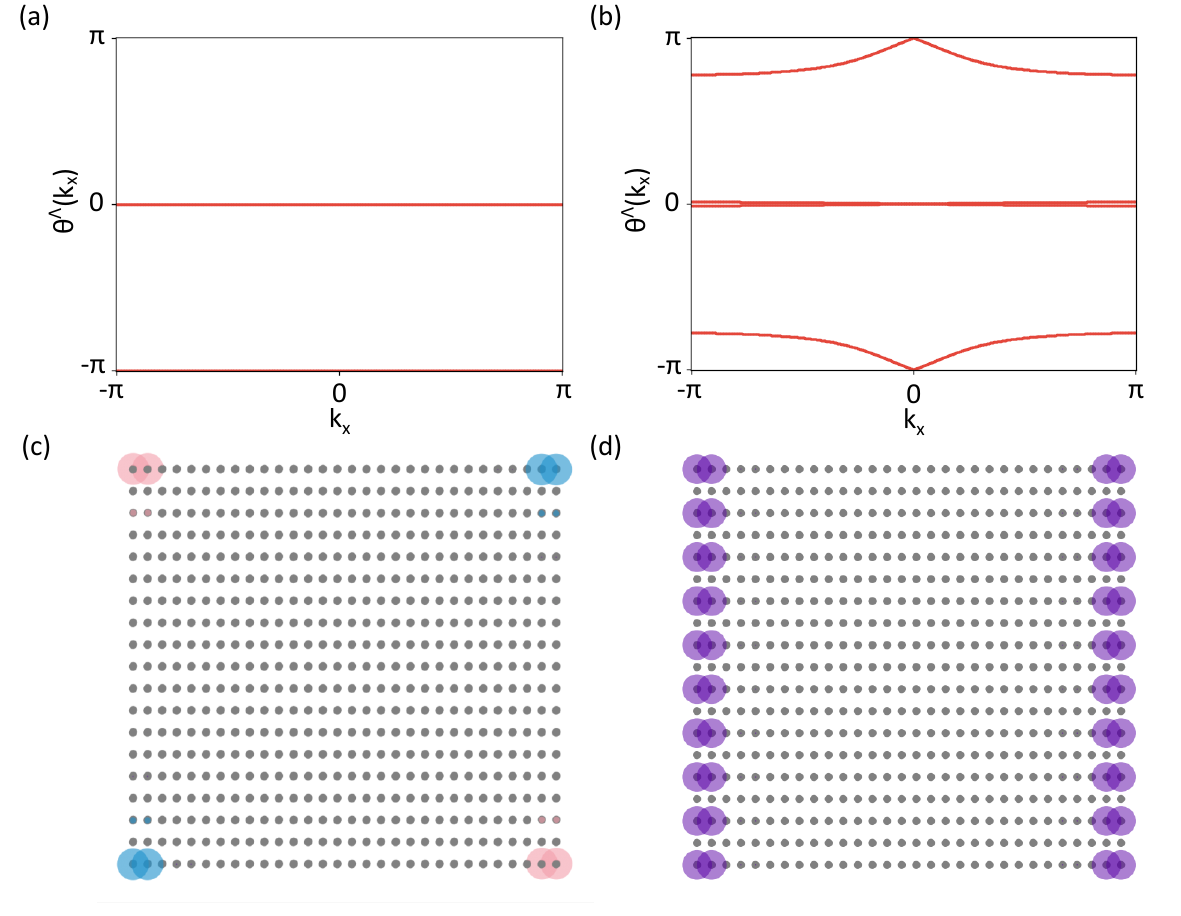}
		\caption{(a) The twisted Wilson-loop spectral flow for the Hamiltonian in  Eq.~\eqref{H3}, which shows a nontrivial 1D invariant. The parameters are set as $t_x^1=1, t_x^2=5, t_y^1=2, t_y^2=4.5$. (b) The twisted Wilson-loop spectral flow for the auxiliary  Hamiltonian in Eq.~\eqref{directsum_3}. The parameters $t_x^1, t_x^2, t_y^1, t_y^2$ are the same in (a), and the parameters of the perturbation are given in Eq.~\eqref{perturbation}. (c) The second-order topological phases for a rectangular-shaped sample with $20\times 19$ sites. A $PT$-related pair of zero-energy corner modes are marked by pink (blue) dots on the diagonal (off-diagonal) direction when $t_y^1>t_y^2$($t_y^1<t_y^2$). (d) $PT$-related first-order helical edge states when $t_y^1=t_y^2$.}
		\label{wilsonloop_3}
	\end{figure}
	
	In this appendix, we consider a 2D topological insulator with noncentered $PT$ symmetry, which is shown in Fig.~\ref{model_3}. Its primitive cell contains six sites and the inversion center is on one of the sites (indicated in the figure). Hence, this model cannot have a primitive cell whose center coincides with the inversion center. In other words, the $PT$ symmetry here must be non-centered. The inversion symmetry is represented as
	\begin{equation}\label{inversion-1}
		\hat{P}=\begin{bmatrix}
			1&0\\
			0&e^{ik_y}
		\end{bmatrix}\otimes\begin{bmatrix}
			0&0&1\\
			0&1&0\\
			1&0&0
		\end{bmatrix}\hat{I}.
	\end{equation}

	Combined with the time reversal symmetry $ \hat{T}=\K\hat{I} $, the spacetime inversion symmetry is represented as
	\begin{equation}\label{spacetime-inversion}
		\hat{P}\hat{T}=\begin{bmatrix}
			1&0\\
			0&e^{ik_y}
		\end{bmatrix}\otimes\begin{bmatrix}
			0&0&1\\
			0&1&0\\
			1&0&0
		\end{bmatrix}\K,
	\end{equation}
	which satisfies
	\begin{equation}
		(\hat{P}\hat{T})^2=1.
	\end{equation}
	We observe that the spacetime inversion operator $\hat{P}\hat{T} $ is $ \k $-dependent.
	
	The tight-binding Hamiltonian is given by 
	\begin{equation}\label{hamiltonian-noncentered}
		\begin{split}
			\H(\k)=&\begin{bmatrix}
				0&t_x^1 & t_x^2e^{-ik_x}&T_1&0&0\\
				t_x^1 &0&t_x^1&0&0&0\\
				t_x^2e^{ik_x}&t_x^1&0&0&0&T_2\\
				T_1^*& 0&0&0&t_x^1&t_x^2e^{-ik_x}\\
				0&0&0&t_x^1&0&t_x^1\\
				0& 0&T_2^*&t_x^2e^{ik_x} &t_x^1&0
			\end{bmatrix}\\
		\end{split},
	\end{equation}
	witn $T_1=-t_y^2+t_y^1e^{-ik_y}$, $T_2=t_y^1-t_y^2e^{-ik_y}$
	
	Since the $PT$ symmetry is non-centered with $\k$-dependence, we have to adopt the twisted Wilson loop method to calculate the Stiefel-Whitney numbers. By the unitary transformation  
	\begin{equation}
		V(k_y)=
	\begin{bmatrix}
		1&0\\0&e^{-ik_y/2}\\
	\end{bmatrix}\otimes \begin{bmatrix}
		-\frac{i}{\sqrt{2}}&0&\frac{i}{\sqrt{2}}\\
		\frac{1}{\sqrt{2}}&0&\frac{1}{\sqrt{2}}\\
		0&1&0
	\end{bmatrix},
\end{equation}
the $ PT $ operator in Eq.~\eqref{spacetime-inversion} can be transformed to
\begin{equation}
	\tilde{P}\tilde{T}=V(k_y)\hat{P}\hat{T}V^{\dagger}(k_y)=\K.
\end{equation}
The Hamiltonian in 
Eq.~\eqref{hamiltonian-noncentered} is transformed by $V(k_y)$ to
\begin{equation}\label{H3}
	\tilde{\H}(\k)=\begin{bmatrix}
		-Q_1&-Q_2& 0&S_1&-S_2&0\\
		-Q_2 &Q_1&\sqrt{2}t_x^1&S_2&S_1&0\\
		0&\sqrt{2}t_x^1&0&0&0&0\\
		S_1&S_2&0&-Q_1&-Q_2&0\\
		-S_2&	S_1&0&-Q_2&Q_1&\sqrt{2}t_x^1\\
		0& 0&0&0&\sqrt{2}t_x^1&0
	\end{bmatrix},\end{equation}
with $Q_1=t_x^2\cos(k_x)$, $Q_2=t_x^2\sin(k_x)$, $S_1=(t_y^1-t_y^2)\cos(k_y/2)$, $S_2=(t_y^1+t_y^2)\sin(k_y/2)$.	Since $V(k_y)$ is not $2\pi$-periodic, the real Hamiltonian $\tilde{\H}(k_x,k_y)$  is not $2\pi$-periodic in $k_y$, but satisfying
\begin{equation}
	\tilde{H}(k_x,k_y+2\pi)=\Lambda\tilde{H}(k_x,k_y)\Lambda^{\dagger},
\end{equation}
where $\Lambda$ is given by
\begin{equation}
	\Lambda=V(k_y+2\pi)V^{\dagger}(k_y)=\begin{bmatrix}
		1&0\\
		0&-1
	\end{bmatrix}
	\otimes\begin{bmatrix}
		1&0&0\\
		0&1&0\\
		0&0&1
	\end{bmatrix}.
\end{equation}

We plot the spectral flow of the twisted Wilson loops of $k_y$-subsystems along  $k_x$ in Fig.~\ref{wilsonloop_3}(a) with $t_x^1=1,t_x^2=5,t_y^1=2,t_y^2=4.5$. Note that this model has six bands, and the two lower bands are chosen as valence bands. We observe that the 1D topological invariant is nontrivial. To obtain $w_2$, we construct the auxiliary Hamiltonian
\begin{equation}\label{directsum_3}
	\mathbf{H}(k_x,k_y)=\begin{bmatrix}
		\tilde{H}(k_x,k_y)&\Delta\\
		\Delta^{\dagger}&\tilde{H}(0,-k_y)\\
	\end{bmatrix}.
\end{equation}
with the perturbation matrix
\begin{equation}\label{perturbation}
	\Delta=\begin{bmatrix}
		1&0\\
		0&0
	\end{bmatrix}\otimes\begin{bmatrix}
		10&0&0\\
		0&10&5\sqrt{2}\\
		0&\sqrt{2}&6
	\end{bmatrix}+\begin{bmatrix}
		0&0\\
		0&1
	\end{bmatrix}\otimes\begin{bmatrix}
		-1/2&0&0\\
		0&3/2&0\\
		0&0&-1
	\end{bmatrix}.
\end{equation} 
The corresponding twisted boundary condition is given as 
\begin{equation}
	\mathbf{\Lambda}=\begin{bmatrix}
		1&0\\
		0&1\\
	\end{bmatrix}\otimes\begin{bmatrix}
		1&0\\
		0&-1
	\end{bmatrix}
	\otimes\begin{bmatrix}
		1&0&0\\
		0&1&0\\
		0&0&1
	\end{bmatrix},
\end{equation}
and the corresponding spacetime inversion operator is represented as $\mathbf{PT}=\K$, which is $ \k $-independent.

The spectral flow of the twisted Wilson loop for  $\mathbf{H}(k_x,k_y)$ is plotted in Fig.~\ref{wilsonloop_3}(b), from which we see that the 2D topological invariant $w_2$ is nontrivial. The corner states corresponding to this nontrivial $w_2$ are shown in Fig.~\ref{wilsonloop_3}(c). Particularly, two second-order boundary mode configurations featuring a pair of $PT$-related corner states. Moreover, we can perform a symmetry-preserving deformation without closing the bulk energy gap, so that the system is transformed to a first-order topological phase with a pair of $PT$-related helical edge states as the critical state between the two second-order phases [Fig.~\ref{wilsonloop_3}(d)].

It is noteworthy that even for lattices with unit-cell center being the inversion center, our method is still useful. This is because such a unit cell may be not natural or not compatible with the boundary conditions. 
In order to study bulk-boundary correspondence, the choice of primitive cell cannot be arbitrary and has to be compatible with boundary conditions, and such a compatible choice has no guarantee of centered $PT$. Our proposed solution is essential for handling such cases as well. 

\section{ 2D topological insulator with centered $PT$ symmetry}\label{B}
\setcounter{figure}{0}
\begin{figure}
	\centering	\includegraphics[scale=0.3]{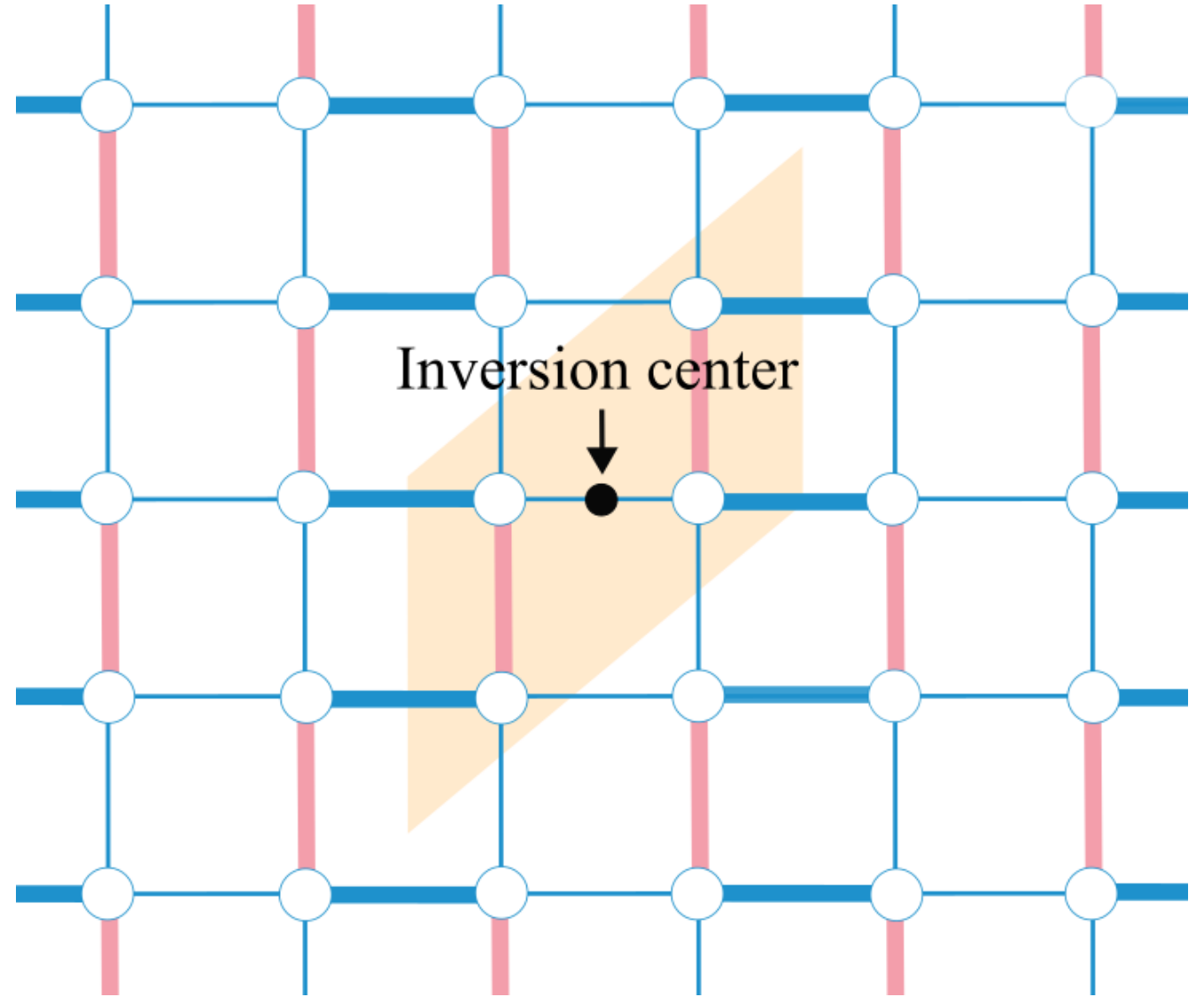}
	\caption{A 2D model on a rectangular lattice. The pink bonds and blue bonds represent negative and positive real hopping amplitudes, respectively. The thick and thin bonds have different hopping strengths. The unit cell is indicated by the green lines, which has centered $PT$ symmetry. }
	\label{model_2}
\end{figure}
\begin{figure}[h]
	\centering	\includegraphics[scale=0.72]{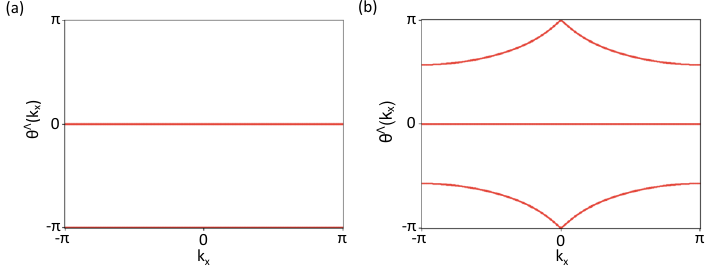}
	\caption{(a) The Wilson-loop spectral flow for the Hamiltonian in  Eq.~\eqref{H2}, which shows a nontrivial 1D invariant. The parameters are set as $t_x^1=1, t_x^2=5, t_y^1=2, t_y^2=4.5$. (b) The Wilson-loop spectral flow for the auxiliary Hamiltonian in Eq.~\eqref{directsum_2}, from which we can extract  $w_2$ for the original model.  The parameters are set as $t_x^1=1, t_x^2=5, t_y^1=2, t_y^2=4.5, t_{01}=t_{13}=2.4$. }
	\label{wilsonloop_2}
\end{figure}

Our method to eliminate nontrivial $1$D topological invariants generally holds regardless of whether the protecting $PT$ symmetry is centered or non-centered. In this appendix, we provide a $2$D model with centered $PT$ symmetry to demonstrate our method. The model is illustrated in Fig.~\ref{model_2}. The unit cell is marked by the shadowed parallelogram, and contains four lattice points. The inversion center is marked by a solid black ball, from which one can easily see that the inversion center is also the unit-cell center. The inversion symmetry is  represented as
\begin{equation}\label{inversion-1}
	\hat{P}=\tau_1\otimes\sigma_0\hat{I}.
\end{equation}
Combined with the time reversal symmetry $ \hat{T}=\K\hat{I} $, the spacetime inversion symmetry is represented as
\begin{equation}\label{spacetime-inversion-2}
	\hat{P}\hat{T}=\tau_1\otimes\sigma_0\K,
\end{equation}
which satisfies
\begin{equation}
	(\hat{P}\hat{T})^2=1.
\end{equation}
Indeed, the spacetime inversion operator  $ \hat{P}\hat{T} $ is $ \k $-independent.

The tight-binding Hamiltonian is given by 
\begin{equation}\label{H2}
	\H(\k)=\sum_{i=1}^{4}f_i(\k)\Gamma^i+g_1(\k)i\Gamma^4\Gamma^5+g_2(\k)i\Gamma^3\Gamma^5.
\end{equation}
Here, the coefficient functions are given by $f_1=-t_y^2+t_y^1\cos k_y$, $f_2=-t_y^1\sin k_y$, $f_3=(t_x^2/2)\sin k_x-(t_x^1/2)\sin k_y-(t_x^2/2)\sin (k_x+k_y)$, $f_4=-(t_x^2/2)\cos k_x+(t_x^1/2)(\cos k_y-1)+(t_x^2/2)\cos (k_x+k_y)$, $g_1=(t_x^2/2)\sin k_x+(t_x^1/2)\sin k_y+(t_x^2/2)\sin (k_x+k_y)$, and $g_2=(t_x^2/2)\cos k_x+(t_x^1/2)(\cos k_y+1)+(t_x^2/2)\cos (k_x+k_y)$, where the $t$'s are hopping amplitudes as indicated in Fig.~\ref{model_2}. The Dirac matrices are chosen as $\Gamma^1=\tau_0\otimes\sigma_1$, $\Gamma^2=\tau_3\otimes\sigma_2$, $\Gamma^3=\tau_2\otimes\sigma_3$, $\Gamma^4=\tau_1\otimes\sigma_3$, and $\Gamma^5=-\tau_3\otimes\sigma_3$, which satisfy $\{\Gamma^\mu,\Gamma^\nu\}=2\delta^{\mu\nu}1_4$.

In Fig.~\ref{wilsonloop_2}(a), we plot the spectral flow of the ordinary Wilson loops of $k_y$-subsystems along $k_x$ for the Hamiltonian in Eq.~\eqref{H2}. The parameters are set as $t_x^1=1, t_x^2=5, t_y^1=2, t_y^2=4.5$, and the number of occupied bands is taken as 2. We can see $w_{1,y}=1$. Thus, one cannot obtain the 2D Stiefel-Whitney number  $w_2$ by conventional method.

According to our method, we can eliminate the nontrivial  $w_{1,y}$ by constructing the Hamiltonian
\begin{equation}\label{directsum_2}
	\mathbf{H}(k_x,k_y)=\begin{bmatrix}
		\H(k_x,k_y)&\Delta\\
		\Delta^{\dagger}&\H(0,-k_y)\\
	\end{bmatrix},
\end{equation}
which preserves the 2D topological invariant $w_2$ of the original model. Here, the mixing perturbation term can be chosen as
\begin{equation}
	\Delta=t_{01}\tau_0\otimes\sigma_1+t_{13}\tau_1\otimes\sigma_3.
\end{equation}

We show the spectral flow of the Wilson loop for $\tilde{H}(k_x,k_y)$ in Fig.~\ref{wilsonloop_2}(b) with $t_x^1, t_x^2, t_y^1, t_y^2$ being the same as the original model and $t_{01}=t_{13}=2.4$. One can readily observe that $w_2$ is nontrivial.

In fact, the generality of our approach stems from its K-theoretical foundation, namely the stability of K-theoretical groups, which is certainly independent of the type of 1D Stiefel-Whitney number. We have demonstrated the general applicability of our method by models in both cases of centered and non-centered $PT$ symmetries.

\section{3D second-order real nodal-line semimetal} \label{C}
In this appendix, we consider a 3D graphite lattice model with noncentered $PT$ symmetry, which was constructed in Ref.~\cite{Shao_2021prl} and is shown in Fig.~\ref{3d_g_d}(a). There is $\pi$ flux for each rectangle and no flux for each hexagon. The lattice has an off-centered inversion symmetry, which is the combination of the screw rotation $ S_\pi $ and the reflection $ M_z $. The screw rotation $ S_\pi $ is projectively represented as
\begin{equation}\label{screw-rotation}
	\S_\pi=\begin{bmatrix}
		0 & 1 \\ e^{ik_z} & 0
	\end{bmatrix}\otimes \sigma_1\hat{I}_{xy},
\end{equation}
with $ \hat{I}_{xy} $ inversing the momentum of $ x-y $ plane. $ M_z $ is represented as
\begin{equation}
	\hat{M}_z=\tau_1\otimes\sigma_0\hat{I}_z.
\end{equation}
Hence, the inversion symmetry is projectively represented as
\begin{equation}\label{inversion-graphene}
	\hat{P}=\begin{bmatrix}
		1 & 0 \\ 0 & e^{ik_z}
	\end{bmatrix}\otimes\sigma_1\hat{I}.
\end{equation}
However,we can recover it by the gauge transformation $\mathit{G}$ that reverses the sign of each site in the even rows, which is represented as
\begin{equation}\mathit{G}=\tau_3\otimes\sigma_0.
\end{equation}
So we are led to the $\mathit{G}$-dressed off-centered inversion symmetry $\hat{P}$ as
\begin{equation}\label{inversion-graphene}
	\hat{P}=\begin{bmatrix}
		1&0\\0&-e^{ik_z}
	\end{bmatrix}\otimes\sigma_1\hat{I}.
\end{equation}
By the time reversal symmetry $ \hat{T}=\K\hat{I} $, the spacetime inversion symmetry is represented as
\begin{equation}\label{spacetime-inversion-1}
	\hat{P}\hat{T}=\begin{bmatrix}
		1 & 0 \\ 0 & -e^{ik_z}
	\end{bmatrix}\otimes\sigma_1\K,
\end{equation}
which satisfies
\begin{equation}
	(\hat{P}\hat{T})^2=1.
\end{equation}

By the unitary transformation  
\begin{equation}\label{unitary-transformation}
	V(k_z)=\frac{1}{2}\begin{bmatrix}
		1&0\\
		0&ie^{-ik_z/2}\\
	\end{bmatrix}\otimes\begin{bmatrix}
		1+i&1-i\\
		1-i&1+i
	\end{bmatrix},
\end{equation}
the $ PT $ operator in Eq.~\eqref{spacetime-inversion-1} can be transformed to
\begin{equation}
	\tilde{P}\tilde{T}=V(k_z)\hat{P}\hat{T}V^\dagger(k_z)=\K,
\end{equation}
The constraint of $ \tilde{P}\tilde{T} $ is also $ \k $-independent.
\setcounter{figure}{0}
\begin{figure}[t]
	\centering
	\includegraphics[scale=0.38]{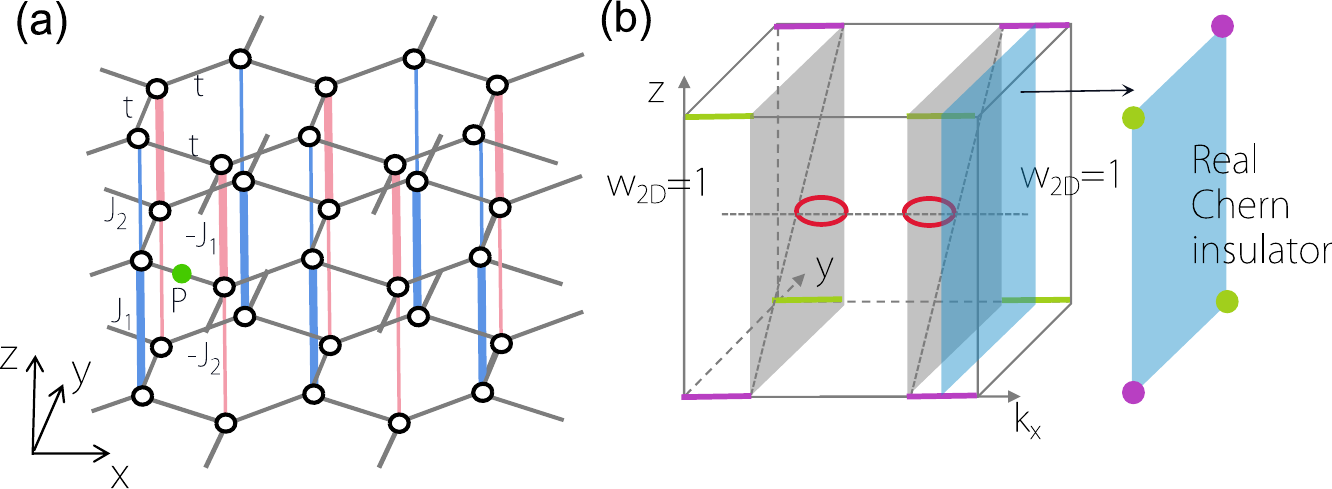}
	\caption{(a) 3D graphite lattice model with flux $\pi$ for each rectangle and no flux for each hexagon. Pink (blue) color marks bonds with a negative (positive) hopping amplitude. Green point indicates the noncentered space inversion. (b) Schematic figure for the phase of nodal loop. Here, the 2D subsystem (blue plane) has a nontrivial second-order SW number $w_2^{(\Lambda)}$ with two pair of corner state related by the inversion symmetry. These 2D corner states form the hinge Fermi arcs connecting the nodal loops. The position of hinge arcs depends on the sign of $ J_- $. }
	\label{3d_g_d}
\end{figure}

\begin{figure}
	\centering	\includegraphics[scale=0.44]{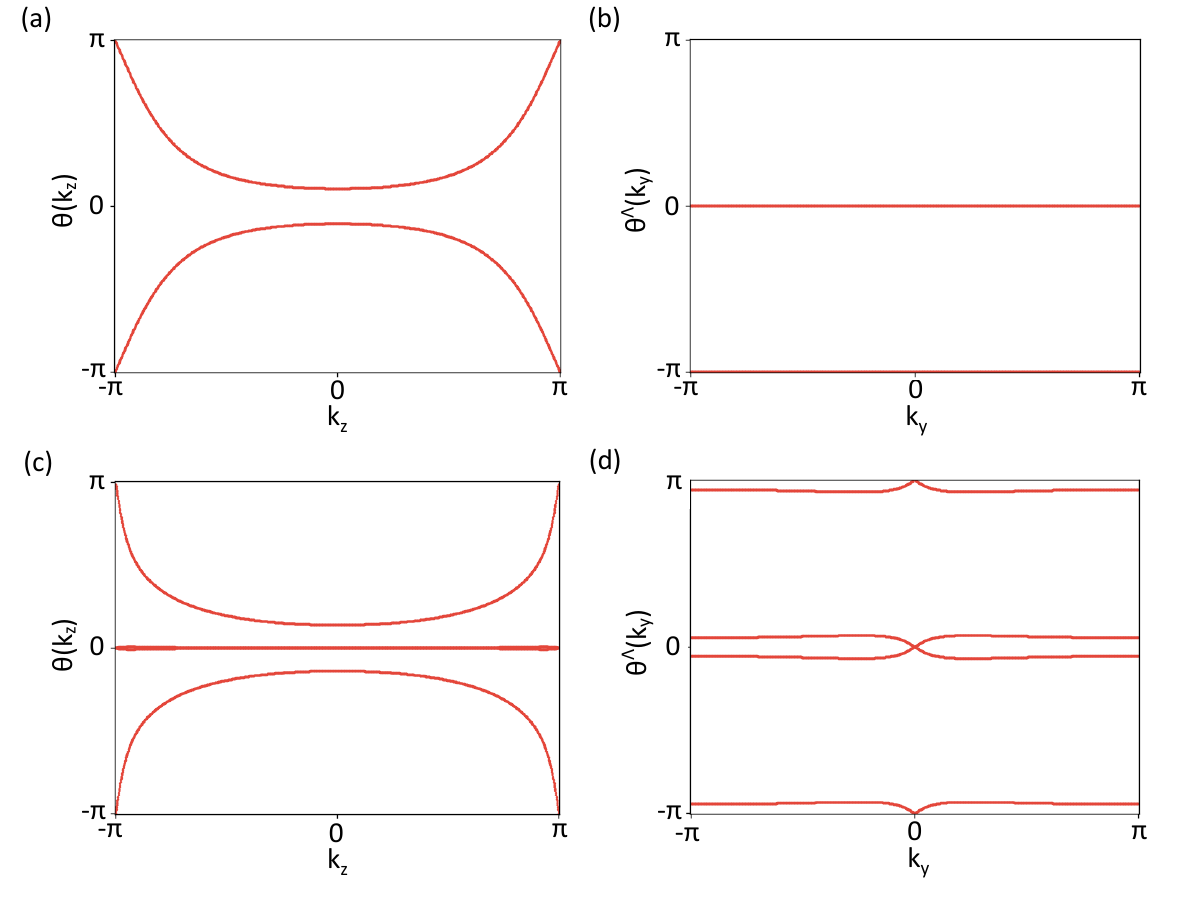}
	\caption{(a) and (b) plot eigenvalues of the twisted Wilson loops along $ k_y $ and $ k_z $ for 2D $k_y-k_z$ subsystems of Eq.~\eqref{2DHam} with $ k_x=\pi $, respectively. The parameters are set as $t=1,J_1=J_2=1$. (c) and (d) plot the eigenvalues for the twisted Wilson loops along $ k_y $ and $ k_z $ for the Hamiltonian in Eq.~\eqref{directsum_hamil} with $ t=J_2=2 $, $ J_1 =1.0$, and $ t_{01}=1.5 $, respectively. }
	\label{wilsonloop_graphite}
\end{figure}
The tight-binding Hamiltonian is given by 
\begin{equation}\label{hamiltonian-graphene}
	\H(\k)=\sum_{i=1}^{4}\chi_i(\k)\Gamma_i+g_1(k_z)i\Gamma^5\Gamma^4+g_2(k_z)i\Gamma^5\Gamma^3,
\end{equation} 
where $\chi_1(\k)+i\chi_2(\k)=t\sum_{i=1}^{3}e^{-i\k\cdot\bm{a}_i}$ with $\bm{a}_i$ the three bond vectors for each hexgonal layer, $\chi_3(\k)=J_+(1+\cos k_z)$, $\chi_4(\k)=J_+\sin k_z$, $g_1=J_-(1-\cos k_z)$ and $g_2=J_-\sin k_z$ with $J_+=(J_1+J_2)/2$ and $J_-=(J_1-J_2)/2$. The Dirac matrices are given as $\Gamma^1=\tau_0\otimes\sigma_1$, $\Gamma^2=\tau_0\otimes\sigma_2$, $\Gamma^3=\tau_1\otimes\sigma_3$, $\Gamma^4=\tau_2\otimes\sigma_3$, and $\Gamma^5=\tau_3\otimes\sigma_3$, which satisfy $\{\Gamma^\mu,\Gamma^\nu\}=2\delta^{\mu\nu}$. By the unitary transformation in Eq.~\eqref{unitary-transformation}, the Hamiltonian in 
Eq.~\eqref{hamiltonian-graphene} can be transformed to
\begin{equation}
	\tilde{H}(\k)=\tilde{f}_1(\k)\Gamma^1+\tilde{f}_2(\k)i\Gamma^1\Gamma^2+\tilde{f}_3(k_z)i\Gamma^5\Gamma^4+\tilde{f}_4(k_z)i\Gamma^5\Gamma^3\Gamma^2,
\end{equation}
where $\tilde{f}_1(\k)=t\sum_{i=1}^{3}\cos(\k\cdot\bm{a}_i)$, $\tilde{f}_2(\k)=t\sum_{i=1}^{3}\sin(\k\cdot\bm{a}_i)$, $\tilde{f}_3(k_z)=(J_1-J_2)\sin(k_z/2)$ and $\tilde{f}_4(k_z)=(J_1+J_2)\cos(k_z/2)$. In Ref.~\cite{Shao_2021prl}, a real Dirac semimetal is realized without the dimerization, $ i.e. $, when $J_-=0$. For $ J_-\neq0 $, the Dirac points are resolved into real nodal loops in presence of the dimerization along the z-direction. The nodal loops have two topological numbers as the first and second Stiefel-Whitney(SW) invariants. One of them is inherited from the real Dirac point, and the other is the Berry phase along the circle encircling the loop. They give rise to hinge helical modes as shown in Fig.~\ref{3d_g_d}(b). The 2D $k_y$-$k_z$ subsystem $\H_{2D}(k_y,k_z)$ parameterized by $k_x$ in the interval connecting the two nodal lines is a 2D SW insulator. When $k_x=\pi$, the Hamiltonian of the subsystem is written as 
\begin{equation}\label{2DHam}
	\tilde{\H}_{2D}(k_y,k_z)=\tilde{\H}(k_x=\pi,k_y,k_z).
\end{equation}
The twisted boundary condition along the $z$ direction is given by  
\begin{equation}\label{twist_boundary}
	\tilde{\H}_{2D}(k_y,k_z+2\pi)=\Lambda\tilde{\H}_{2D}(k_y,k_z)\Lambda^{\dagger},
\end{equation}
where  $\Lambda$ is given by 
\begin{equation}\label{lambda}
	\Lambda=V(k_z+2\pi)V^{\dagger}(k_z)=\tau_3\otimes\sigma_0.
\end{equation}
According to Eq.~\eqref{w1}, one can get $w_1^{(\Lambda)}(k_z)=0$ and $w_1^{(\Lambda)}(k_y)=1$ as shown in Fig.~\ref{wilsonloop_graphite}(a) and \ref{wilsonloop_graphite}(b) which plot the eigenvalues of the twisted Wilson loop for Eq.~\eqref{2DHam}. Fig.\ref{wilsonloop_graphite}(b) implies the first SW invariant depends on the direction of Wilson loop, which signifies the existence of weak topology. To stripping off weak topology, we follow the procedure of the main text by making a direct sum of a $\tilde{H}_{2D}(k_y,k_z)$ and $\tilde{H}_{2D}(0,-k_z)$, and adding some perturbations $\Delta$ such that
\begin{equation}
	[\tilde{P}\tilde{T},\ \Delta]=0,[\Lambda,\ \Delta]=0.
\end{equation} 
By this procedure, the corresponding spacetime inversion operator is represented as
$\mathbf{PT}=\K$,
whose constraint is $ \k $-independent too. To be precise, we have the resultant Hamiltonian as
\begin{equation}\label{directsum_hamil}
	\mathbf{H}(k_y,k_z)=\begin{bmatrix}
		\tilde{\H}_{2D}(k_y,k_z)&\Delta\\
		\Delta^{\dagger}&\tilde{\H}_{2D}(0,-k_z)\\
	\end{bmatrix},
\end{equation}
and the corresponding twisted boundary condition is given as
\begin{equation}
	\mathbf{\Lambda}=\begin{bmatrix}
		\Lambda&0\\
		0&\Lambda\\
	\end{bmatrix},
\end{equation}
with $ \Lambda $ given in Eq.~\eqref{lambda}. After introducing the perturbation $ \Delta=t_{01}\tau_0\otimes\sigma_1 $, we calculate the eigenvalues of the twisted Wilson loops for the resulting Hamiltonian in Eq.~\eqref{directsum_hamil} as shown in Fig.~\ref{wilsonloop_graphite}(c) and \ref{wilsonloop_graphite}(d). Compared with Fig.~\ref{wilsonloop_graphite}(b), the strong topology
is resolved from the weak topology along the $k_z$ direction with $ w^{(\Lambda)}_2=1 $, $ i.e. $, the spectra of Wilson loops cross $ \pi $ an odd number of times.



\bibliographystyle{apsrev4-2}
\bibliography{kDepen_Ref}

\begin{thebibliography}{42}
\bibitem [{Inv()}]{Inversion_Center_Notes}%
  \BibitemOpen
  \href@noop {} {}\bibinfo {note} {The unit cell is determined by the
  translational symmetry. Hence, inversion symmetry does not necessarily
  preserve the integarity of each unit cell, i.e., it can map sites in a unit
  cell to sites belonging to more than one unit cells. In this case, the
  inversion is called noncentered, and is momentum dependent under the
  unit-cell convention.}\BibitemShut {Stop}%
\bibitem [{Not()}]{Note_Components}%
  \BibitemOpen
  \href@noop {} {}\bibinfo {note} {Since $D(k_x) \in O(\mathcal{N})$, the
  eigenvalues $e^{i\theta_a}$ must either equal $\pm 1$ or form
  complex-conjugate pairs. $O(\mathcal{N})$ consists of two disconnected
  components, $O(\mathcal{N})=SO(\mathcal{N})\sqcup MSO(\mathcal{N})$, where
  $M$ is a mirror operator reversing one coordinate of the N dimensional space.
  $w_{1,y} = \pm 1$ dictates whether $D(k_x)$ lies in $SO(\mathcal{N})$ or
  $MSO(\mathcal{N})$.}\BibitemShut {Stop}%
\bibitem [{Z2_()}]{Z2_Notes}%
  \BibitemOpen
  \href@noop {} {}\bibinfo {note} {For $\mathcal{N}$ = 2, the classiffcation is
  $\pi_1(SO(2)) = \mathbb{Z}$. If we add one more trivial band, only the parity
  of the integer is stable when the trivial band is mixed with other bands,
  i.e., $\mathbb{Z}$ is reduced to $\mathbb{Z}_2$.}\BibitemShut {Stop}%
\bibitem [{Pat()}]{Patch_Notes}%
  \BibitemOpen
  \href@noop {} {}\bibinfo {note} {To circumvent the problem, one might want to
  transform $D(k_x)\in MSO(\mathcal{N})$ by
  $\tilde{D}(k_x)=[D(-\pi)]^{-1}D(k_x)$. Then, $D(k_x)$ is topologically
  nontrivial if and only if $\tilde{D}(k_x)$ is. The latter can be checked by
  inspecting the spectral flow since $\tilde{D}(k_x)\in SO(\mathcal{N})$. But,
  in numerical calculations, the \textit{uncontrolled} $U(k_x)$ obstructs such
  a transformation, noting that the trivial trial $[W(-\pi)]^{-1}W(k_x)$ is not
  an operator equivalent to $SO(\mathcal{N})$ up to a unitary
  transformation.}\BibitemShut {Stop}%
\bibitem [{Eve()}]{EvenCell_Notes}%
  \BibitemOpen
  \href@noop {} {}\bibinfo {note} {Proof: Suppose one can choose a primitive
  cell for which the inversion center is also the unit-cell center, then sites
  in the unit cell form inversion-related pairs except the one at the center.
  This leads to an odd number of sites, which is a contradiction}\BibitemShut
  {NoStop}%
\end{thebibliography}

\end{document}